\documentclass[prl,nofootinbib,twocolumn]{revtex4}

\usepackage{amsmath}
\usepackage{amsfonts}
\usepackage{graphicx}

\begin{document}

\title{String-inspired ultraviolet/infrared mixing
and preliminary evidence\\
of a violation of the de Broglie relation for nonrelativistic neutrons}

\author{Giovanni AMELINO-CAMELIA and Flavio MERCATI}
\affiliation{{\small Dipartimento di Fisica, Universit\`a di Roma ``La Sapienza"}
{\small and Sez.~Roma1 INFN, P.le Moro 2, Roma, Italy}\\
{\small $~~~~~~~~~$ amelino@roma1.infn.it, flavio.mercati@gmail.com $~~~~~~~~~$}}

\begin{abstract}
We advocate a novel perspective on the phenomenology of
a framework with spacetime noncommutativity which is of
established relevance for string theory. Our analysis applies to cases
in which the noncommutativity parameters are arranged
according to the criteria of ``light-like noncommutativity"
and ultraviolet supersymmetry is assumed,
so that the implications of the characteristic mechanism of ultraviolet/infrared
mixing are relatively soft.
We also observe that an analogous case of soft ultraviolet-infrared mixing is present
in a previously-proposed Loop-quantum-gravity-inspired description of quantum spacetime.
And we show that soft ultraviolet-infrared mixing produces an anomaly
for the nonrelativistic de Broglie relation $\lambda v = h/m$, with correction term governed
by a single (but particle-dependent)
parameter $\chi$. We test this hypothesis by comparing
a determination
of the fine
structure constant that relies on the de Broglie relation for nonrelativistic neutrons
to other independent determinations of the fine structure constant,
and we obtain an
 estimate of $\chi$ that differs from 0 with
four-standard-deviation significance.

%
%
%
%
%
%
%
%
%
%
%
\end{abstract}

\maketitle

In the long, and so far inconclusive, hunt for
a quantum theory of gravity the main strategy
was inspired by the discovery paradigm
of the 20th century, the ``microscope paradigm"
with discovery potential measured in terms of
the shortness of the distance scales probed.
But it is actually expected that quantum gravity,
besides certainly providing new phenomena in a
far-UV (ultraviolet) regime, should also have
significant implications in a dual IR (infrared)
regime. For example, our present understanding
of black-hole thermodynamics, and particularly
the scaling $S \propto R^2$ of the entropy
of a black hole of radius $R$,
suggests that such IR features are inevitable~\cite{cohenUVIR,joseUVIR}.
It is therefore perhaps no accident that
noteworthy IR features were indeed found~\cite{szabo}
in the study of the only
theories in a ``quantum spacetime" that proved
to be manageable with presently-available methods
of computation and interpretation,
which is the case of theories with ``canonical"
noncommutativity of spacetime coordinates,
  $[x_\mu ,x_\nu]=i \theta_{\mu \nu}$.
 Intriguingly it has also been robustly established that
this type of spacetime noncommutativity emerges in various
regimes~\cite{szabo} of String Theory, the most studied candidate for quantum gravity.
For example, a mechanism analogous to the emergence of noncommutativity
of position coordinates
in the Landau model also characterizes the
description of strings in presence of a constant
Neveu-Schwarz two-form B-field~\cite{szabo}.

We shall here focus on models
with ``light-like" noncommutativity matrix~\cite{lightconeNCFT1,lightconeASCHIERI}
($\theta_{\mu\nu} \theta^{\mu\nu} = \epsilon_{\mu\nu\rho\sigma} \theta^{\mu\nu} \theta^{\rho\sigma}=0$)
and assuming UV supersymmetry.
The results of analyses of self-energy corrections~\cite{szabo,susskind}
 in such light-like-$\theta_{\mu\nu}$
 scenarios can motivate the study of  modifications of the on-shell
relation of the form
\begin{eqnarray}
m^2 \simeq E^2 - p^2 + \chi_\theta \, m^2 \log \left( \frac{E+{\vec{p}} \cdot {\hat{u}}_\theta}{m} \right)
~.
 \label{softIR}
\end{eqnarray}
The parameter $\chi_\theta$ takes different values for different particles~\cite{szabo,susskind}
and the  unit vector ${\hat{u}}_\theta$ describes a preferential
direction~\cite{lightconeNCFT1}
determined by the matrix $\theta_{\mu \nu}$.

Our interest in the specific scenario (\ref{softIR}) is in part
motivated by the fact that it appears to characterize a type
of IR behaviour which might also emerge in other approaches
to quantum-gravity/quantum-spacetime research. This is suggested
by the quantum-spacetime model of Refs.~\cite{urrutiaPRL,urrutiaPRD},
further developed\footnote{In this first exploratory study we shall,
like Refs.\cite{josePRD,gacFlavioPRL2009},
set aside the possibility of an helicity dependence~\cite{urrutiaPRD} of the effects.}
in parts of Refs.\cite{josePRD,gacFlavioPRL2009},
which was inspired by one of the competing perspectives on
the semi-classical limit of Loop Quantum Gravity,
and found corrections  to the dispersion relation with dominant
IR/long-wavelength behaviour linear in momentum~\cite{urrutiaPRL,urrutiaPRD}.
Clearly also (\ref{softIR}) produces effects whose dominant IR  behaviour
is linear in momentum, since for small $p$ ($E\simeq m$) one
finds $\log [(E+{\vec{p}} \cdot {\hat{u}}_\theta)/m]
 \propto {\vec{p}} \cdot {\hat{u}}_\theta$. The long-wavelength behaviour of the two models
 therefore differs only because of the fact that invariance under spatial rotations
 (lost in (\ref{softIR})) is preserved by the scenario of
 Refs.~\cite{urrutiaPRL,urrutiaPRD,josePRD}.
In most of the following we shall consider simultaneously the two models,
by observing
that the characterization of (\ref{softIR}) in terms of $\chi_\theta$
and ${\hat{u}}_{\theta}$ is applicable, in the long-wavelength regime,
to the scenario of Refs.~\cite{urrutiaPRL,urrutiaPRD,josePRD}
by replacing ${\hat{u}}_{\theta}$ with ${\hat p} \equiv {\vec{p}}/p $
(which correctly reproduces the space-rotation invariance assumed
in Refs.~\cite{urrutiaPRL,urrutiaPRD,josePRD})
and replacing $\chi_\theta$ with an independent parameter $\chi_{\hat p}$,
since any given set of data will constrain differently
 the strength of the effect in the two models (as a result of their different
 description of the fate of space-rotation symmetry).

While the merit of preserving space-rotation symmetry may appear
to be a decisive argument in favour of the $\chi_{\hat p},{\hat p}$ setup,
we should stress that the case of Eq.~(\ref{softIR}) has the priority
for what concerns the robustness
of the link to a candidate quantum-gravity theory,
since its  relevance for string theory has been investigated
in detail~\cite{szabo,lightconeNCFT1}.
From that perspective it is perhaps useful for us to comment
on alternatives to the choice of light-like $\theta_{\mu \nu}$
and UV supersymmetry.
 If the model completely lacks supersymmetry
(even in the UV limit) then the IR features can have inverse-power-law dependence~\cite{susskind}
 on momentum, producing much more virulent IR effects than expected with (\ref{softIR}).
 And if the $\theta_{\mu \nu}$
  matrix is not of light-like type,
 one ends up either spoiling unitarity~\cite{szabo,lightconeNCFT1} (``time-like noncommutativity") or producing
 IR features that exclusively involve the spatial components of momentum
 (``space-like noncommutativity"), in ways that
 produce a singular long-wavelength limit~\cite{szabo,susskind}.

In the very extensive (mostly string-inspired) literature
devoted to spacetime noncommutativity,
 the presence of IR corrections, and the ``UV/IR-mixing mechanism"~\cite{szabo,susskind}
that produces them,
have attracted very strong interest. However, they were
mainly contemplated as technically intriguing
but physically cumbersome features.
We shall here adopt a complementary perspective,
perceiving their possible use as explicit
examples of IR properties for the quantum-gravity
realm as the key reason of interest in these models.

Consistently with the intuition we have been
developing in an ongoing investigation of broader scopes~\cite{tesiPHDflavio},
our characterization of the long-wavelength properties of theories
in quantum spacetime focuses on modifications of the ``nonrelativistic
de Broglie relation" (whose standard form is $\lambda v = h/m$).
We here derive such modifications assuming
an on-shell relation
of the form $m^2 \simeq E^2 - p^2 + \chi \, m \, {\vec{p}} \cdot {\hat{u}}$,
which for $\chi = \chi_\theta$, ${\hat{u}} = {\hat{u}}_\theta$
describes the case of the long-wavelength
limit of (\ref{softIR}), while
for $\chi = \chi_{\hat p}$, ${\hat{u}} = {\hat p}$ describes
the case of
the long-walength properties of the model of
Refs.~\cite{urrutiaPRL,urrutiaPRD,josePRD}.
From such linear-in-momentum modifications of the on-shell relation
one derives a modified nonrelativistic de Broglie relation
which takes the form
\begin{eqnarray}
\lambda {\vec v}  \simeq \frac{ h}{m} {\hat{p}}
-  \chi \frac{\lambda}{2} \hat{u}  +O\left(\lambda v^3\right)
~,
 \label{dbA}
\end{eqnarray}
where we assumed that $v^j =\partial E/\partial p_j$ still holds.

By focusing on the nonrelativistic de Broglie relation
the relevant phenomenology has the advantage,
as we shall discuss in more detail elsewhere~\cite{tesiPHDflavio},
of probing a
rather robust feature of the type of models we
 are considering. In fact,
 modifications to $\lambda v = h/m$ are found both if $v(p)$ is anomalous
 ($v \neq p/(p^2+m^2)^{1/2}$)
and if $v(p)$ is undeformed but there are deformations
of the wavelength-momentum relation ($\lambda \neq h/p$,
a modified ``relativistic de Broglie relation").
 Let us here illustrate this briefly
 within the example of the result of Refs.~\cite{urrutiaPRL,urrutiaPRD},
 which was obtained by analyzing propagation of waves in the relevant quantum-spacetime picture,
 and therefore is most faithfully described in terms of a
 dispersion relation
  $4 \pi^2 m^2/h^2 \simeq \omega^2 - k^2 + \chi_{\hat p} \, 2 \pi \, m \, k/h$.
From this one obtains
Eq.~(\ref{dbA}) assuming undeformed wavelength-momentum
relation
 ($k \equiv 2 \pi/\lambda = 2 \pi p/h$), and thereby obtaining from the group velocity, $\partial \omega / \partial k$,
 an anomalous dependence of velocity on
 momentum: $\vec v \simeq h  \vec k/(2 \pi m) - \chi_{\hat p} \, \hat k/2 \simeq
 \vec p/m - \chi_{\hat p} \, \hat p/2$.
However, in the spirit of  Ref.~\cite{kowaDB,kempf,ahluDB} and references therein,
some authors assume that in such cases the relationship between velocity and momentum
should still not be anomalous,
advocating a consistency requirement between modifications of the $\omega(k)$
dispersion relation and modifications of the
wavelength-momentum relation.
In particular, for the model of Refs.~\cite{urrutiaPRL,urrutiaPRD},
following the scheme of Ref.~\cite{kowaDB},
one could assume $\vec k \simeq 2 \pi \vec p/h + 2 \pi \chi_{\hat p} m {\hat p}/(2 h)$, which produces
and undeformed velocity-momentum relation:
 $\vec v \simeq h  \vec k/(2 \pi m) - \chi_{\hat p} \, \hat k/2 \simeq
 \vec p/m$. However, one then also finds  $\lambda \equiv 2 \pi/k \simeq h /(p+\chi m/2)$
 which combines with $\vec v \simeq  \vec p/m$
 to produce once again (\ref{dbA}).

An unpleasant aspect of the phenomenology we are proposing originates from
the severe particle dependence of $\chi$ that one expects from the theory side,
which limits our ability to identify a preferred type of particle probes.
For the model based on spacetime noncommutativity, here parametrized
with $\chi_\theta$, this particle dependence is derived rigorously~\cite{susskind}
and is rather virulent: for a given quantum field the coefficient $\chi_\theta$
depends on the strength and number~\cite{szabo,susskind} of its interactions with other fields, including
interactions with possible ultramassive fields
(fields  on which
our current particle-physics
laboratories can provide no information, but here relevant because of the UV/IR mixing~\cite{susskind}).
Moreover, according to some perspectives on the quantum-gravity problem
there might even be additional sources of particle dependence, linked to the compositeness of
particles:
various semi-heuristic arguments suggest~\cite{qgPARTONS1,qgPARTONS2} that
 the short distance structure of
spacetime should affect ``composite particles" (clearly atoms, and perhaps
even hadrons within a parton picture) in a way that gets softer
for higher compositeness because of a sort of averaging-out/coarse-graining effect.
We shall not dwell on these hypotheses here, but they did tentatively direct toward hadrons
and leptons our search of examples of measurements that could be used to illustrate
our proposal.

This literature search quickly informed us of the fact
that, outside the quantum-gravity community,
 the de Broglie relation is by now an unquestioned cornerstone
of the laws of physics.
We found no reasonably recent review
of tests of de Broglie relation.
Even precision experiments
assume it to be exactly valid, and often the analysis is reported
in such a way that a reader can no longer
disentangle the information on the role played
by the de Broglie relation. Among the few exceptions to this (in our opinion unhealthy)
state of affairs the most precise measurement we could find with an intelligible role
played by the nonrelativistic de Broglie relation is the one of Ref.~\cite{kruger1995e1998}.
This concerns a study of neutrons
which determined
 their wavelength $\lambda$,
in terms of the $d_{220}$ lattice spacing~\cite{martinBecker} of high-perfection silicon crystals,
 and also their speed ($v \simeq 1600~\text{m/s}$).
The final result can be given in the form~\cite{kruger1999}
\begin{eqnarray}
 \frac{\lambda v}{d_{220[W04]}} = 2\,060.267\,004(76) ~ \text{m/s}
~,
 \label{krugeRES}
\end{eqnarray}
where $d_{220[W04]}$ is the $d_{220}$
lattice spacing of the silicon crystal WASO04~\cite{martinBecker,kruger1999},
and numbers in parentheses are one-standard-deviation uncertainties in the last digits.

We observe that one can accurately test the nonrelativistic
de Broglie relation for neutrons, $\lambda v=h/m_n$,
by comparing the result (\ref{krugeRES})
to experimental determinations of $h/(m_n d_{220[W04]})$
based on the formula~\cite{kruger1995e1998}
\begin{eqnarray}
 \frac{h/m_n}{d_{220[W04]}}
  =  \frac{1}{d_{220[W04]}} \,\, \alpha^2 \frac{m_e}{2 R_\infty m_n}
~,
 \label{hmalpha}
\end{eqnarray}
where both $m_e/m_n$  (ratio of
electron and neutron mass) and the
Rydberg constant $R_\infty$
($R_\infty \equiv \alpha^2 m_e/(2 h)$)
are very accurately known experimentally~\cite{kruger1995e1998}.
Through precision measurements of ${d_{220[W04]}}$
and of the fine structure constant $\alpha$
one can determine  $h/(m_n d_{220[W04]})$ using (\ref{hmalpha}) and check
the agreement with
the result (\ref{krugeRES}), required for the validity
of the de Broglie relation. This in turn would allow to deduce
bounds\footnote{Note that for particles of speed $v \simeq 1600~\text{m/s}$
one can rely on (\ref{dbA}) (applicable
for $v^3 \ll |\chi|$) only to probe $|\chi|$
not smaller than $\sim \!\! 10^{-15}$. To probe even smaller
 values of $|\chi|$ one should either include the leading relativistic correction
 or use slower particles.}
on $\chi$,
in the same spirit of other ``quantum-gravity phenomenology"
studies~\cite{grbgac,schaeferPRL,gacQM100}.

Of course, in Refs.~\cite{kruger1995e1998,kruger1999},
and in papers that followed, the exact validity of the nonrelativistic
de Broglie relation is assumed {\it a priori},
and the result (\ref{krugeRES}) is mainly used for a determination of the fine structure
constant (exploiting the relationship (\ref{hmalpha})), often denoted with $\alpha_{h/m_n}$.
The relatively recent Refs.~\cite{harvardPRL2006,codata,birabenEPJ2008} did
compare $\alpha_{h/m_n}$ to other independent determinations of $\alpha$,
but their assessment of the situation was in part
affected\footnote{We gratefully acknowledge generous feedback
by Francois Biraben and Francois Nez,
which helped us becoming comfortable with our understanding of the
limitations of the characterization of the comparison of $\alpha_{h/m_n}$ and $\alpha$
found
in Refs.~\cite{harvardPRL2006,birabenEPJ2008,codata}.}
by the related errata in Refs.~\cite{erratumCAVAGNERO,erratumGABI}.
Both in (Fig.1 of) Ref.~\cite{harvardPRL2006}
and in (Fig.1 of) Ref.~\cite{birabenEPJ2008}
the $\alpha_{h/m_n}$ result
of Ref.~\cite{kruger1999} was described as fully
consistent with the most accurate available
determinations of $\alpha$,
but that comparison was implicitly affected by ignoring the erratum
published in Ref.~\cite{erratumCAVAGNERO} which invalidates previous
understandings of $d_{220[W04]}$.
On the other hand, (page 1248
of) Ref.~\cite{codata} pointed
out a disagreement of $2.8$-standard-deviation significance between
$\alpha_{h/m_n}$ and independent measurements of $\alpha$,
but a key item for that analysis, concerning determinations of $\alpha$ based on
measurements of the electron  $g\!-\!2$,
could not take into account the relevant erratum in Ref.~\cite{erratumGABI}.

For our assessment of the  implications of the
results for $\lambda v/d_{220[W04]}$ of
Ref.~\cite{kruger1999}
we rely on two recent results,
which allow a very accurate determination of $h/(m_n d_{220[W04]})$.
For $\alpha$ we rely on Ref.~\cite{gab08}, which reported a remarkably accurate
determination, $\alpha^{-1} \! = \! 137.035\,999\,084(51)$, obtained combining a
measurement of the electron $g\!-\!2$ and an elaborate field-theory computation
 of the relationship between $\alpha$ and the electron $g\!-\!2$
within QED. The error in the QED computation that was reported in
Ref.~\cite{erratumGABI} has been resolved (the result is now fully confirmed by
two independent derivations~\cite{gab08}) and the experimental uncertainty
of the measurement of  the electron $g\!-\!2$ has been reduced
by a factor 3 with respect to the best previous measurements.
For $d_{220[W04]}$ we rely on the very recent Ref.~\cite{hallelu2009},
which reported $d_{220[W04]} = 192\,015. 570\,2(10)~\text{fm}$, a four-fold
improvement in accuracy with respect to the best previous determinations.

Both of these improved measurements affect the analysis in the direction
of a wider mismatch between $h/(m_n d_{220[W04]})$
and the result for $\lambda v/d_{220[W04]}$ obtained in Ref.~\cite{kruger1999}.
As illustrated in Fig.~1, previous
determinations of $h/(m_n d_{220[W04]})$ were already higher
than $\lambda v/d_{220[W04]}$ of Ref.~\cite{kruger1999}.
The recent precise determination of $\alpha$ reported in Ref.~\cite{gab08}
is
(moderately) higher than the
value of $\alpha$ assumed in Ref.~\cite{harvardPRL2006,codata}, and this in turn
produces a higher estimate of $h/(m_n d_{220[W04]})$.
And the recent precise measurement of $d_{220[W04]}$ reported in Ref.~\cite{hallelu2009}
is noticeably smaller than the previous corresponding ``world average",
a reduction which also goes in the direction of further increasing the
estimate of $h/(m_n d_{220[W04]})$.
The net result is that our analysis exposes a four-standard-deviation discrepancy:
combining the result (\ref{krugeRES}) of Ref.~\cite{kruger1999},
the determination of $\alpha$  recently reported in Ref.~\cite{gab08},
and the measurement of $d_{220[W04]}$  recently reported in Ref.~\cite{hallelu2009},
we find $(h/m_n - \lambda v)/(h/m_n) = (1.46 \pm 0.37) \cdot 10^{-7}$.
For the $\chi_{{\hat p}[n]}$ parameter ($\chi_{\hat p}$ for neutrons)
this would imply
$$\chi_{{\hat p}[n]} = (1.55 \pm 0.39) \cdot 10^{-12}~.$$

\begin{figure}[hb!]
\includegraphics[width=0.48\textwidth]{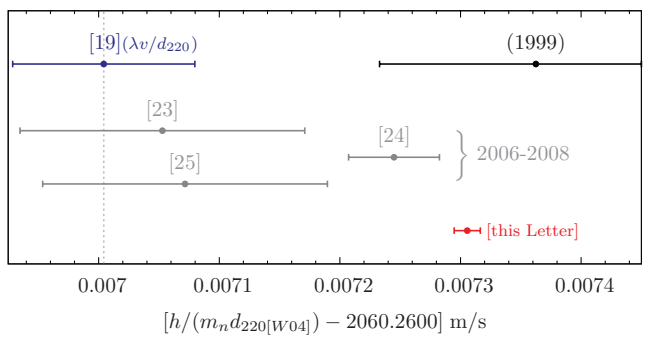}
\caption{We compare the result for $\lambda v/d_{220[W04]}$
of Ref.~\cite{kruger1999} to some estimates of $h/(m_n d_{220[W04]})$
that (explicitly or implicitly) were provided in the
literature. There was reasonably good agreement
in 1999 (publication of \cite{kruger1999}).
From Refs.~\cite{harvardPRL2006}, \cite{codata} and \cite{birabenEPJ2008},
published between 2006 and 2008,
one could obtain different
estimates of  $h/(m_n d_{220[W04]})$,
all affected in part
by the errata in \cite{erratumCAVAGNERO}
and/or \cite{erratumGABI}.
Our updated determination of  $h/(m_n d_{220[W04]})$ simply combines
recent precise measurements
of $d_{220[W04]}$ (Ref.\cite{hallelu2009})
and $\alpha$ (Ref.\cite{gab08})}
\end{figure}

The lack of information on orientation
of the apparatus for the data reported in Ref.~\cite{kruger1999}
does not allow us to derive a definite estimate of $\chi_{\theta [n]}$.
Still, unless one is willing to assume that by chance the data
reported in Ref.~\cite{kruger1999} happened to sample in exactly uniform way all
possible orientations of the neutron velocities (a ``conspiracy' which, with a finite number
of measurements, is even logically problematic),
our findings would also imply, assuming the validity of Eq.~(\ref{softIR}),
that $ \chi_{\theta [n]} \neq 0$ with four-standard-deviation significance,
also setting~\cite{tesiPHDflavio}
a {\underline{lower}} bound: $ \chi_{\theta [n]} > 0.77 \cdot 10^{-12} $.

While it is amusing to offer estimates of $\chi_{{\hat p}[n]}$ and $\chi_{\theta [n]}$
clearly what we exposed here is at best intriguing preliminary evidence of a violation of the
nonrelativistic de Broglie
relation, which would also admit description in models of such violations that are not of
the type~\cite{tesiPHDflavio} we here focused on.
And even the apparent crisis
for the nonrelativistic de Broglie relation should of course be perceived with healthy skepticism,
particularly considering the magnitude of the implications
for our present description of the laws of physics.
Still we do feel that, indeed because of the paramount importance
that such a discovery would carry, this small crisis should be
promptly
investigated experimentally.

It seems that priority should be given to investigations of
the most vulnerable link in our characterization of the discrepancy, which is the fact that
the original result of Refs.~\cite{kruger1995e1998,kruger1999} has not been
checked
in a repetition of the experiment.
The final estimate reported in Ref.~\cite{kruger1999}
conscientiously combined (throughout-consistent) data gathered
over two stages of measurements, separated by a full reinstallation and upgrade
of the apparatus, and for five somewhat different configurations. But with
some of the novel techniques
developed over this past decade (see, {\it e.g.}, Ref.~\cite{hallelu2009} and references therein)
 a significant improvement of the result should be easily within reach.
In light of the particle dependence of the effect
 expected from the theory side, and of the prudence that from a more general
perspective our present (lack of) understanding of the quantum-gravity problem demands,
it would be ideal to have new precision tests of the de Broglie relation again
for neutrons, and for velocities of the neutrons comparable to the ones
produced in the setup of Refs.~\cite{kruger1995e1998,kruger1999}.
It is certainly also interesting to search for analogous effects for atoms,
especially since in some cases
(such as caesium, $\chi_{[Cs]}$, and rubidium,  $\chi_{[Rb]}$)
very small values of $\chi$ could perhaps be probed.
A sizable difference between $\chi_{[n]}$  and $\chi$ of atoms with large
mass number might not be surprising, but
according to our personal theoretical prejudice
relatively small differences should be found between the case of neutrons and
the case of Hydrogen atoms.

Even if, as it is natural to expect, the experimental situation
 eventually settles removing the anomaly here exposed, our
study could still inspire at least a small shift of focus for quantum-gravity theory research:
as here illustrated,
there might be windows of opportunity
for quantum-spacetime phenomenology in precision measurements of properties of the
long-wavelength regime, and perhaps
 a bit more of the theory effort could be diverted from
the ``microscope paradigm" toward targeting
these possible opportunities.

We thank F.~Biraben, J.~Cortes,
G.~Gubitosi, F.~Lacava, C.~L\"ammerzahl,
F.~Nez,
L.~Smolin,
G.~Tino, D.~Villamaina and C.~Voena.

\end{document}